\documentclass[twocolumn]{oja}
\usepackage{graphicx}
\usepackage{amsmath}
\usepackage{enumitem}
\usepackage{orcidlink}

\def\Msolar{\ifmmode {\rm M_{\odot}}\else $\rm M_{\odot}$\fi}
\def\Mearth{\ifmmode {\rm M_{\oplus}}\else $\rm M_{\oplus}$\fi}
\def\Rearth{\ifmmode {\rm R_{\oplus}}\else $\rm R_{\oplus}$\fi}
\def\micron{\ifmmode {\mu{\rm{m}}}\else $\mu$m\fi}

\begin{document}

\renewcommand*{\sectionautorefname}{Section} 
\renewcommand*{\subsectionautorefname}{Section} 
\renewcommand*{\subsubsectionautorefname}{Section} 

\title{Light Echoes of Time-resolved Flares and Application to Kepler Data}

\author{Austin J. King\,\orcidlink{0000-0001-9181-1105}}
\author{Benjamin C. Bromley\,\orcidlink{0000-0001-7558-343X}}
\affil{Department of Physics \& Astronomy, University of Utah, 
\\ 115 S 1400 E, Rm 201, Salt Lake City, UT 84112}
\email{austin.king@utah.edu}


\begin{abstract} 
Light echoes of stellar flares provide an intriguing option for exploring protoplanetary disks in young stellar systems. Previous work on light echoes of circumstellar disks made use of delta-function flares for modeling. We present a new model that incorporates echoes produced by extended, time-resolved flares. We then test this model on known disk-bearing stars with Kepler K2 data by estimating disk parameters from possible echo signals. We focus on two stars; the first appears to be a good candidate for use of this echo model, which predicts disk parameters that are consistent with known values. The second star turns out to be more problematic as a result of high brightness variability in its post-peak lightcurve. These two cases show both the promise and limitations of light echoes as a tool for exploring protoplanetary disks in the time domain.
\end{abstract}

\keywords{Planetary systems --- flare stars --- disks}

\section{Introduction} 
\label{intro}
Protoplanetary disks are the first major structures to take shape around newly formed stars and are the birthplaces of planets \citep{Kenyon08b, youdin2013}. These disks can have radial extents of hundreds of AU \citep{facchini17, Tazaki25}. Protoplanetary disks are not permanent features of their systems --- they decay over time as material accretes into the host star, experiences outflows and photoevaporation, or condenses as gas and dust is swept up by newly formed planetesimals \citep{Williams11}. Stellar mass seems to play a role in the persistence of protoplanetary disks, in addition to age of the system. \cite{Ribas2015} found that for stars of $<2M_\odot$, protoplanetary disks are present in about $60-70\%$ of stellar systems between the ages of 1-3 Myr, and $5-20\%$ between 3-10 Myr. Beyond 10 Myr, about $10\%$ of these low-mass stars still host a disk. Conversely, they noted that only $35-40\%$ of high-mass stars ($>2M_\odot$) host disks between 1-3 Myr, with $0\%$ after 3 Myr. 

Not only are young stars more likely to host protoplanetary disks - they also tend to have high rates of activity, with flare rates and amplitudes decreasing after about 50 Myr \citep{davenport2019, feinstein2020}. This makes young, disk-bearing systems the ideal candidates for testing and application of the light echo detection method.

\begin{figure*}[ht]
    \centering
    \includegraphics[scale=0.6]{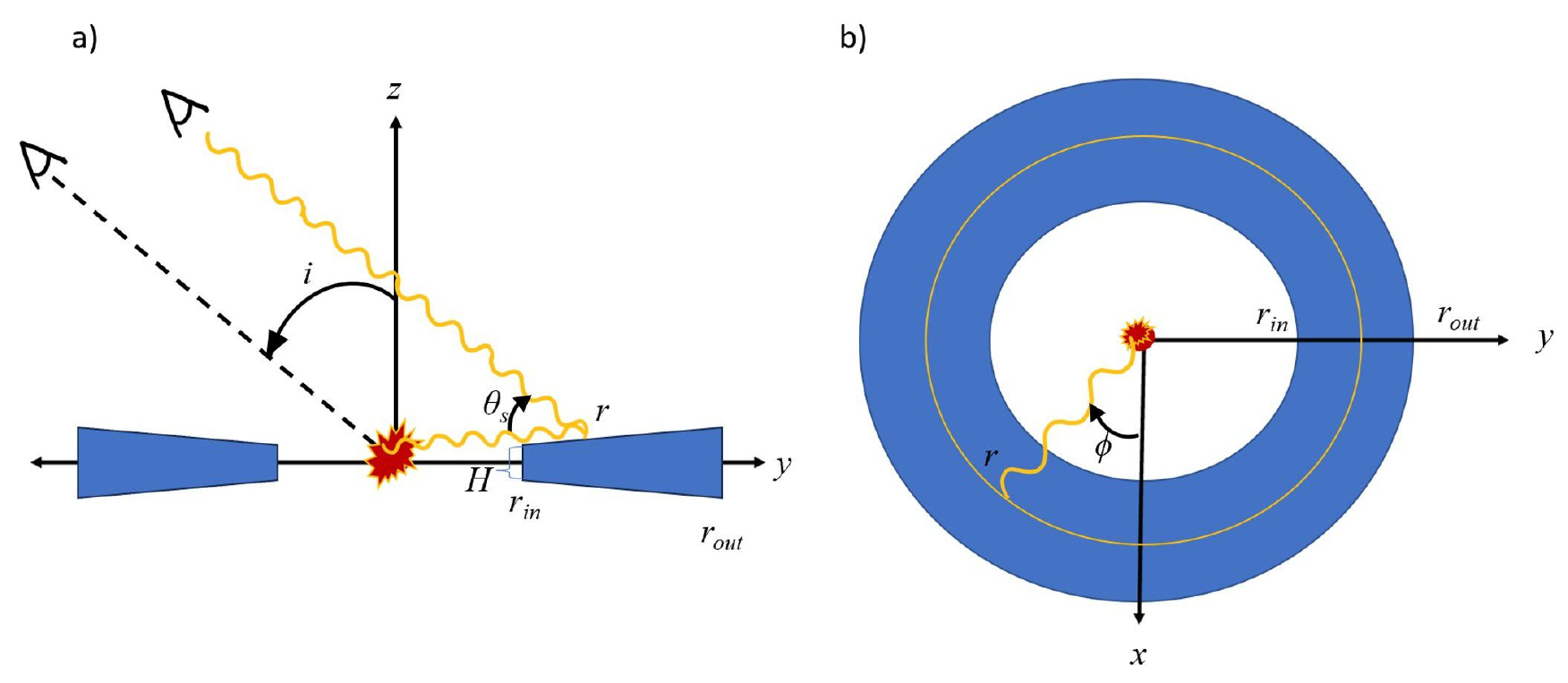}
    \caption{Disk Model: a) A cross-sectional side view along the x-axis of the protoplanetary disk. The flash at the origin depicts a stellar flare. The yellow squiggles show the path of a photon as it strikes a
    point at radius $r$ on the disk and is scattered toward the observer at a scattering angle of $\theta_s$. $i$ denotes the inclination of the disk relative to the observer ($i=0^\circ=$ face-on, $i=90^\circ=$ edge-on), $H$ is the physical thickness of the disk at its inner radius $r_{\rm in}$, and $r_{\rm out}$ is the outer radius. b) a top-down view of the disk along the z-axis. $r_{\rm in}$ and $r_{\rm out}$ once again denote inner and outer radii, respectively. The yellow ring shows all angles, $\phi$, that a photon traveling to distance $r$ will interact with before scattering to the observer. Note: In this orientation, the portion of the disk aligned with the negative y-axis would be defined as the ``near side", as it is inclined toward the observer.}
    \label{fig:geometry}.
\end{figure*}

Following a sudden peak in brightness, light will propagate outward and scatter off surrounding circumstellar material. Recent works exploring how to interpret the resulting light echo have considered models describing the characteristic echo profile produced from a variety of dusty or gaseous disk configurations. These include optically and geometrically thin debris disks \citep{bromley2021, King24} and optically thick, geometrically flared protoplanetary disks \citep{King24-2}. Figure~\ref{fig:geometry}, taken from \cite{King24-2}, illustrates the geometry of a protoplanetary disk interacting with flare light.

Each of these studies advanced the light echo method, building upon the ways that it can be applied. However, all of these previous studies on light echoes of disks assumed that stellar flares are ideal delta functions.  While this assumption is a good approximation to impulsive flares, it cannot be accurately applied to general cases, including large-amplitude flares that tend to be long-lasting \citep[e.g.,][]{hawley2014}. 

The Kepler Space Telescope has proved to be an invaluable resource for collection and study of flare events. Kepler has compiled flare data on M dwarf stars \citep[e.g.]{hawley14,davenport14}, superflare data of solar-type stars \citep{shibayama2013}, and helped develop a catalog of over 4000 flaring stars \cite{davenport2016b}. Kepler data of young stellar systems has been used to determine the presence of disks, presented in \citet{cody18},\citet{venuti21}, and \citet{cody22}. These works are of particular interest for this paper due to the confirmation of disks around active stars, providing a great data set for testing the light echo method. The Transiting Exoplanet Survey Satellite (TESS) could provide useful data for this venture as well. However, it is important to note that TESS focuses on brighter stars than Kepler \citep{ricker2015}. The brightness of these stars may make detection of relatively fainter echoes more difficult. For now, we keep focus on Kepler data to test the capabilities of the light echo method. 

Here, we extend previous work on the detection of echoes from circumstellar disks by including flares that are resolved in the time domain. In the next section (\S\ref{sec:echomod}, we describe echoes of delta-function flares --- the starting point for this work --- and how we make adjustments to handle long-lasting flares. Then, in \S\ref{sec:data}, we apply this model to candidate stars with known protoplanetary disks from Kepler K2 light curve data and present our findings regarding the potential success of this method. We summarize in \S\ref{sec:summary}.

\section{Echo Modeling}\label{sec:echomod}
The structure outlined in \cite{King24-2} can be utilized for modeling echoes protoplanetary/optically thick disks with a flared geometry. In this optically thick condition, the Draine phase function \citep{draine2003}
\begin{equation}\label{eq:draine}
    \Phi_\alpha(\theta_s) = 
    \frac{1}{4\pi} 
    \left[\frac{(1-g^2)}{(1+\alpha(1+2g^2)/3}\right] \frac{1+\alpha\cos^2\theta_s}{(1+g^2-2g\cos{\theta_s})^{3/2}}
\end{equation} 
is calculated with adjustable scattering parameters $g$ and $\alpha$ set as $g=\alpha=0$, which yields isotropic scattering and reduces the function to $1/4\pi$. 

The angular position on the disk, $\phi$, is measured from 0 to $2\pi$. The zero-point is taken to be along the positive x-axis of the disk, shown in part b) of Figure~\ref{fig:geometry}, and the positive y-axis is directed toward the far side of the disk, angled away from the observer. The remaining physical parameters of the disk --- inner radius, $r_{\rm in}$, outer radius, $r_{\rm out}$, inclination, $i$, and albedo, $\gamma$, are left as variables. 

The disk is then divided into area elements of size 
\begin{equation}
    dA(r)=2\pi rdrd\phi\cos(\pi/2-\psi(r))
\end{equation}
, where $\psi(r)$ is the slope of the disk as a function of radius. The cosine term accounts the foreshortening of the scattering surface as seen from the source, which lowers the available incident flux for scattering. At the inner wall of the disk, we calculate the area elements as $dA(r=r_{\rm in})=2\pi r_{\rm in}H_{1/2}d\phi$, where $H_{1/2}$ is half the physical disk thickness.

Because of the flared geometry of the disk, the radial position $r$ between $r_{\rm in}$ and $r_{\rm out}$ is calculated from the scale-height relationship $r=\sqrt{r_{\rm m}^2 + H_{1/2}^2}$. In this relationship, $r_{\rm m}$ denotes the radius along the mid-plane of the disk and $H_{1/2}$ is half the physical thickness of the disk, calculated by $H_{1/2}=(r_{\rm m} h)$. The scale height $h$ is determined based on the cadence of the data being used. Short-cadence data, in which Kepler takes exposures of approximately 2-minute time bins, allows for observations of inner regions of planetary systems where the scale height is relatively constant at about $h=0.045$ out to about 10 AU \citep{Bitsch15}. Long-cadence data ($\sim$30-minute bins) can be used to explore regions beyond 10 AU, where the scale height follows a power-law relationship with radius, defined by \cite{Chiang10} as
\begin{equation}
    h_{\rm C\&Y}=0.022(\frac{r_{\rm m}}{\text{AU}})^{2/7}
\end{equation}
. To compensate for the foreshortening of the disk bins, we opt for an increase of this scale height by a factor of 3 ($h=3h_{\rm C\&Y}$). This allows the model to produce viable echoes and is consistent with heights for positions of settled surface grains \citep{Chiang10}.

We can then  calculate the brightness (flux) of each element via the following equation:
\begin{equation}\label{eq:B}
    B = \gamma\frac{dA}{4\pi r^2}\left|1-\left|\phi/\pi-3/2\right|\right|
\end{equation}
where the absolute value term deals with angle-averaging of flares across multiple events as flares on the side of the star nearest to the observer are more likely to be observed. This also shifts the positions of our maximum and minimum brightnesses to the near and far sides of the disk, respectively. Note that we correct an error by \cite{King24-2} where a $\sin(i)$ scaling was used in the angle-averaging term. Inclusion of this term led to unequal values at $\phi = 0, 2\pi$, which should not be the case.

The brightness from each area element is then paired with a corresponding time bin via the crossing time function, defined by \cite{gaidos1994}:
\citet{gaidos1994}:
\begin{equation}\label{eq:T}
    T=\frac{r}{c}\left(1+\sin{\phi}\sin{i}\right).
\end{equation}

The brightnesses of each area element falling within a shared time bin are summed, resulting in a delay distribution representing the disk impulse response consistent with the disk parameters. 

In previous work which employed this style of disk echo calculation, this process was carried out assuming delta-function flares \citep{bromley2021,King24,King24-2}. For this work, we include an additional step to this process to allow the model to fit extended flare light
curves. We introduce $f(t)$, normalized to a peak of 1, to represent the shape of the time-resolved flare light curve. The echo response of the disk is then a convolution of the flare light curve with the delay distribution obtained by summing over the disk surface elements. In the delta-function case, $f(t)$ reduces to a single value equal to 1. This term is included in the brightness function, so it is now described as 
\begin{equation}\label{eq:B_new}
    B_k = f_k\gamma\frac{dA}{4\pi r^2}\left|1-\left|\phi/\pi-3/2\right|\right|
\end{equation}

\begin{figure*}[ht]
    \centering
    \includegraphics[width=.34\textwidth, height=.1875\textwidth]{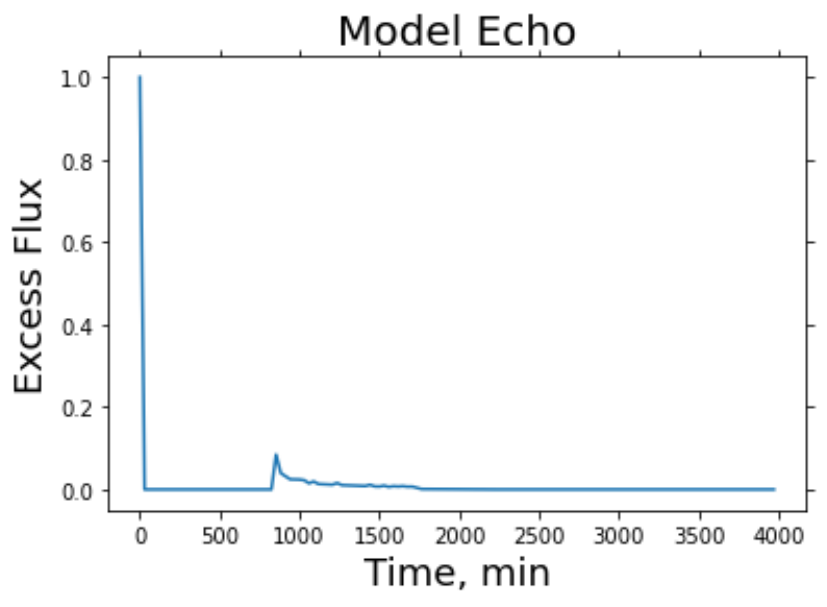}
    \includegraphics[width=.34\textwidth, height=.1875\textwidth]{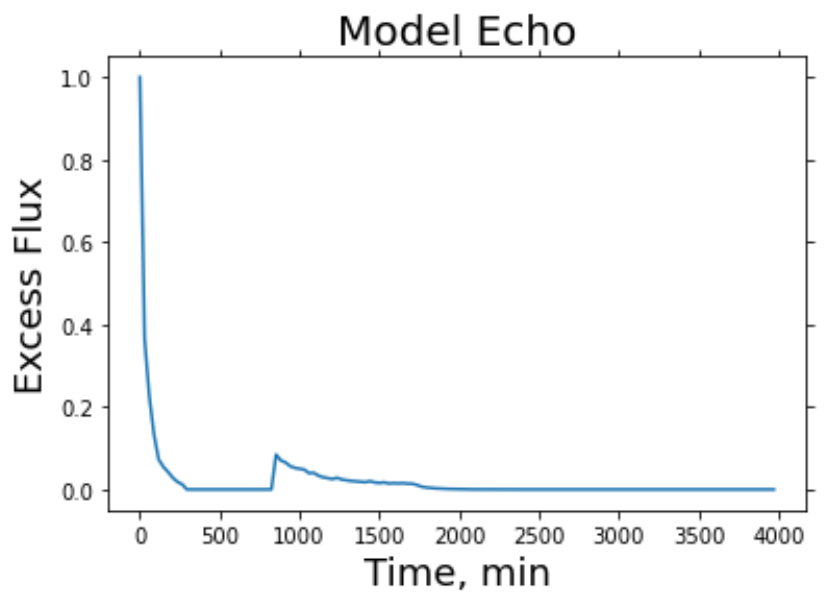}
    \includegraphics[width=.34\textwidth, height=.1875\textwidth]{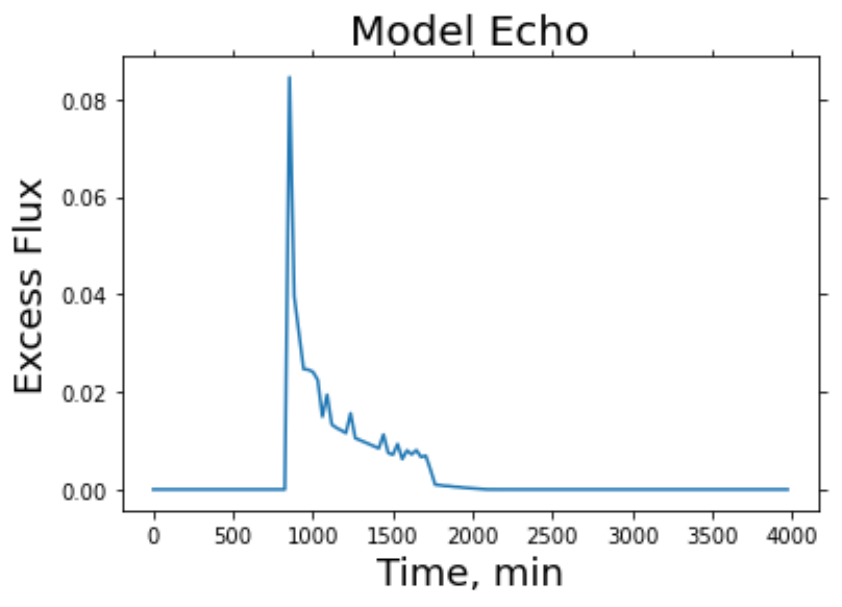}
    \includegraphics[width=.34\textwidth, height=.1875\textwidth]{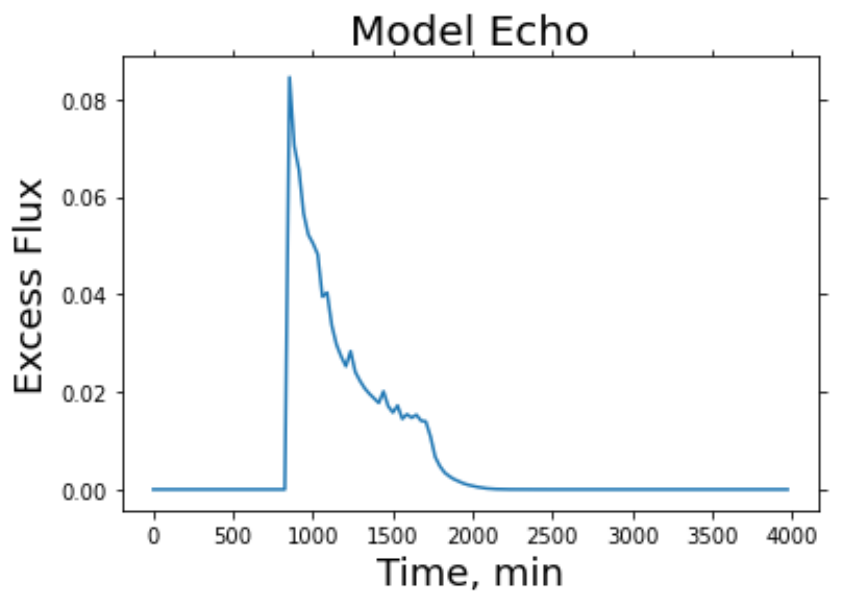}
    \caption{Echo model results for disk of parameters $r_{\rm in}=150\text{ AU, }r_{\rm out}=180\text{ AU, }i=20^\circ \text{, and }\gamma=1$. 
    Left Panels: Delta-function flare with following echo; lower panel scales the lightcurve to show the post-flare echo more clearly.
    Right panels: Time-resolved flare with following echo; lower panel scales the lightcurve to show the post-flare echo more clearly. $\Delta t$ here is approximately 30 minutes, consistent with Kepler long-cadence data.}
    \label{fig:models}
\end{figure*}

In this framework, the brightness $B_k$ is calculated for each value of $f_k$ through the duration of the flare, for $k=0,1,2,...$, where $k$ is the time bin associated with each flare flux value. For each step in $k$, the brightness of elements with shared crossing times are summed, utilizing the numpy function numpy.roll() to shift the calculated light-curve by $k$ time bins, which results in the proper echo delay from each part of the flare.
For $k=0$ and $f_k=1$, a delta-function flare is recovered. We present ideal light curves in Figure~\ref{fig:models}, produced for a disk with the following physical parameters: $r_{\rm in}=150\text{ AU, }r_{\rm out}=180\text{ AU, }i=20^\circ \text{, and }\gamma=1$. The left panels show light curves produced by a single, delta-like impulse with peak value of 1. The top panel shows the full curve including the delta impulse and the bottom panel shows only the section of the curve beginning 500 minutes post-peak. The right panels show the same as the left, but for a more realistic, extended flare. The gradually declining flare results in additional light, driving a brighter and slightly longer lasting echo. The intensity of these curves is described as an "excess flux", where zero excess flux corresponds to zero increase above the quiescent starlight. This is measured in arbitrary units of percent flux increase. This term is used later on in the paper to describe the same property in real data.

We include an additional change here to aid in modeling edge-on or near-edge-on disks. These highly inclined systems should be subject to some degree of obscuration, as light from the near side of the disk is unable to travel through the disk and toward the observer. To account for this, we take a dot product of the unit vector normal to the surface of the disk and the unit vector directed toward the observer and cut where this dot product is less than zero. This results in no light reaching the observer from the near-side of the disk in highly inclined geometries, as is anticipated.

Later, as an additional test when fitting real data, we attempted to include a secondary model that instead treats the circumstellar structure as having a steep, cylindrical wall. We retain inclination and albedo as parameters, but instead vary only one radius, measured to the inner wall of the cylinder, in addition to the vertical thickness of this wall, $H$. Nearly all previous calculations are done the same way, with the exception of our differential area bin size, which is now calculated as $dA=2\pi rd\phi H$. We then set a prior on the height of the cylinder, constraining it between $1$ and $75\%$ of the radius of the cylinder. The high wall geometry does result in more drastic shadowing/self-occultation at high inclinations - this is handled similarly to the shading cut described for the disk, setting the brightness of occulted bins to zero.

\subsection{The impact of extended flares on parameter estimation }\label{subsec:mcmc}
To showcase the improved modeling and parameter extraction abilities of the extended flare light echo model, we estimate disk parameters through a Markov-Chain Monte Carlo (MCMC) fitting process. This follows the steps presented in \cite{King24-2}, beginning with a $\chi^2$ function defined as 
\begin{equation}\label{eq:chi2}
    \chi^2 = \sum_{j=1}^{N}\left(F_{{\rm data}, j} - F_{\rm model}(t_j-t_0, r_{\rm in}, r_{\rm out}, i)\right)^2/\sigma_j^2
\end{equation}
where $t_0$ is the time of flare peak and $j$ denotes time bin index. This function calculates goodness-of-fit for model echoes produced by specific disk parameters to the potential echo in co-aligned post-flare data.

Next, we assume independent Gaussian errors and define our log probability as 
\begin{equation}\label{eq:logP}
    \ln{\mathcal{P}} = -\frac{1}{2}\chi^2 -\frac{1}{2}\sum_{i}^N\ln\left({2\pi\sigma_i^2}\right) + \ln{\mathcal{P}_{\rm priors}}
\end{equation}
where $\mathcal{P}_{\rm priors}$ is defined as
\begin{equation}
    {\cal P}_{\rm priors} \sim  
    \begin{cases}
    \ \sin(i) \ & \ \ 0^\circ \leq i \leq 90^\circ \ \text{and} \\
    & r_{\rm min} \leq r_{\rm in} < r_{\rm out} \leq r_{\rm max}, 
    \\
    0 & \text{otherwise.}
    \end{cases} 
\end{equation}
The prior $\sin(i)$ accounts for randomly oriented disks being more likely to be observed in some approximately edge-on configuration. We apply an uninformative prior on albedo, uniformly distributed between zero and one.

In this mock data, the $\sigma_i$ and time spacing used are taken from the real data presented later in this paper.

\begin{figure*}[!ht]
    \centering
    \includegraphics[width=.7\textwidth, height=.45\textwidth]{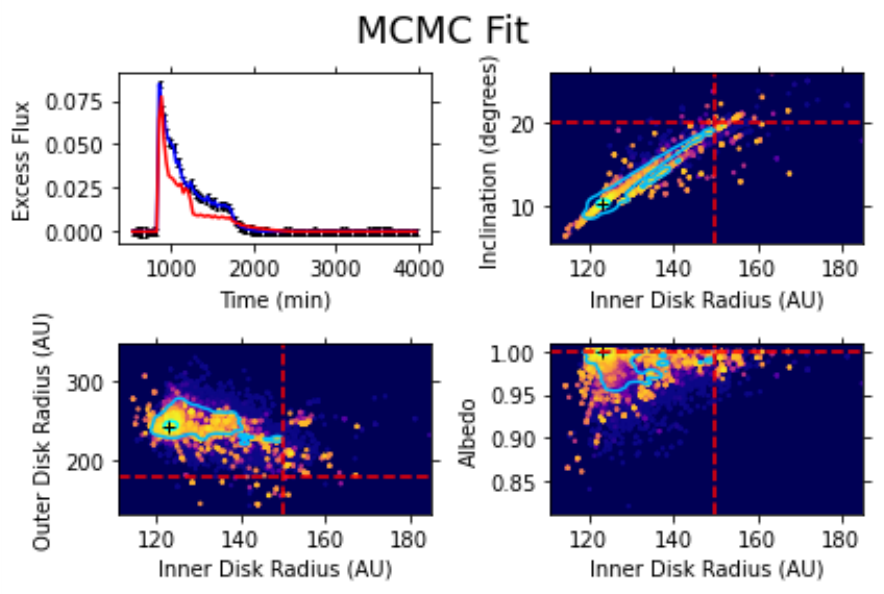}
    
    \caption{MCMC fit with Delta Model:
    Top Left: This panel shows the ideal model echo (red) produced by the maximum a posteriori (MAP) parameters returned from the MCMC fit overlayed on the extended flare echo from Figure~\ref{fig:models} (blue) via a delta flare echo model. Remaining panels each show parameter space fits with blue and cyan contours depicting $1 \sigma$ and $3\sigma$ confidence ranges, respectively. The black plus signs indicate the most probable parameters in each space. The red dashed lines show the true parameter values.}
    \label{fig:delta-fit}
\end{figure*}

\begin{figure*}[!ht]
    \centering
    \includegraphics[width=.7\textwidth, height=.45\textwidth]{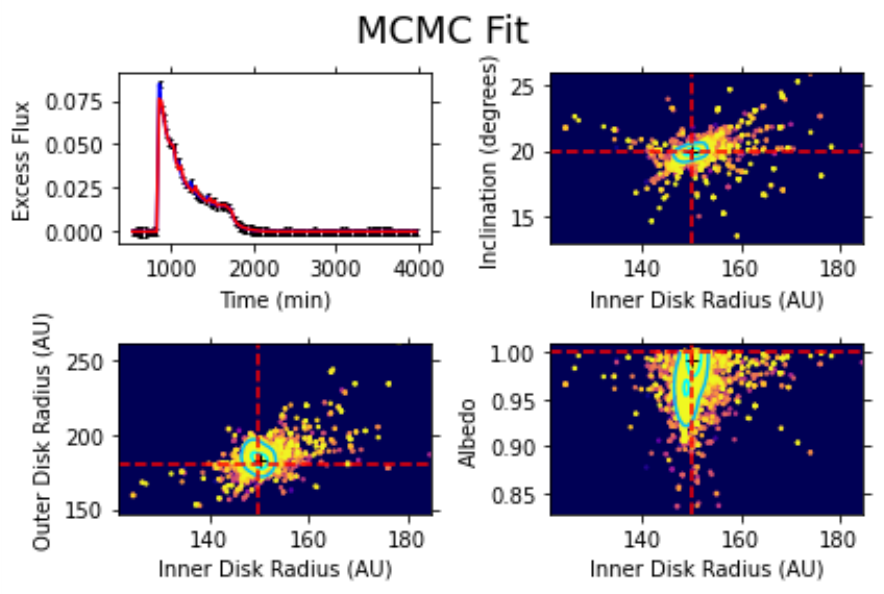}
    
    \caption{MCMC fit with Extended Model:
    Top Left: This panel shows the ideal model echo (red) produced by the MAP parameters returned from the MCMC fit overlayed on the extended flare echo from Figure~\ref{fig:models} (blue) via an extended flare echo model. Remaining panels each show posterior parameter distributions with blue and cyan contours depicting $1 \sigma$ and $3\sigma$ confidence ranges, respectively. The black plus signs indicate the most probable parameters in each space.}
    \label{fig:extended-fit}
\end{figure*}

With these likelihood and goodness-of-fit functions in place, we employ the emcee MCMC python package \citep{emcee2012}. A twenty-step burn-in stage is run before resetting the sampler. After this has been completed, the sampler is given 600 walkers and 4000 steps, surpassing the autocorrelation time for the disk parameters. We complete this process for delta and extended flares, both of which are used in an attempt to fit the echo produced by the extended flare shown in Figure~\ref{fig:models}. The results are shown in Figure~\ref{fig:delta-fit} for the delta-function flare model
and in Figure~\ref{fig:extended-fit} for the extended flare model.

We note that only the results presented in Figure~\ref{fig:extended-fit} correctly return the parameters used to produce the extended flare echo in Figure~\ref{fig:models}, with the exception of albedo. The results of the delta function flare modeling overestimate the outer radius of the disk and underestimate the inclination and inner radius. This confirms that the extended flare echo model in this case is necessary for accurately fitting potential echoes produced by realistic, time-resolved flares.

With the updated echo model in place, we look toward applying it to Kepler data.

\section{The Data}\label{sec:data}
To further test the capabilities of the extended light echo model for protoplanetary disks, we utilize Kepler K2 (K2YSO) data. This survey collected long-cadence (~30-minute exposure) data on 475 young, disk-bearing stars in Upper Scorpius and Ophiuchus \citep{cody18}, the Lagoon nebula \citep{venuti21}, and Taurus \citep{cody22}. We remove modulations in the data via the LightKurve smoothing package \textit{flatten} \citep{LK}, which employs a Savitsky-Golay smoothing algorithm. We use the default values for this function of windowsize=101 and polyorder=2. We tested this function by piping through the theoretical curves presented in Section~\ref{sec:echomod}. We found that we were able to return the same echo signature without degradation and parameter fitting was not significantly affected and the model successfully returned the original input parameters.

After surveying the data, we chose to present two candidates from this catalog for the purposes of this paper: EPIC 203664569, a rotating variable star in Upper Scorpius, is found to be a suitable candidate for application of the light echo method after summing its observed flare events; for comparison, we present EPIC 247584113, an Orion variable in Taurus, which does not appear to be a suitable candidate for further light echo studies.

\subsection{EPIC 203664569}

\begin{figure*}[!ht]
    \centering
    \includegraphics[width=.34\textwidth, height=.1875\textwidth]{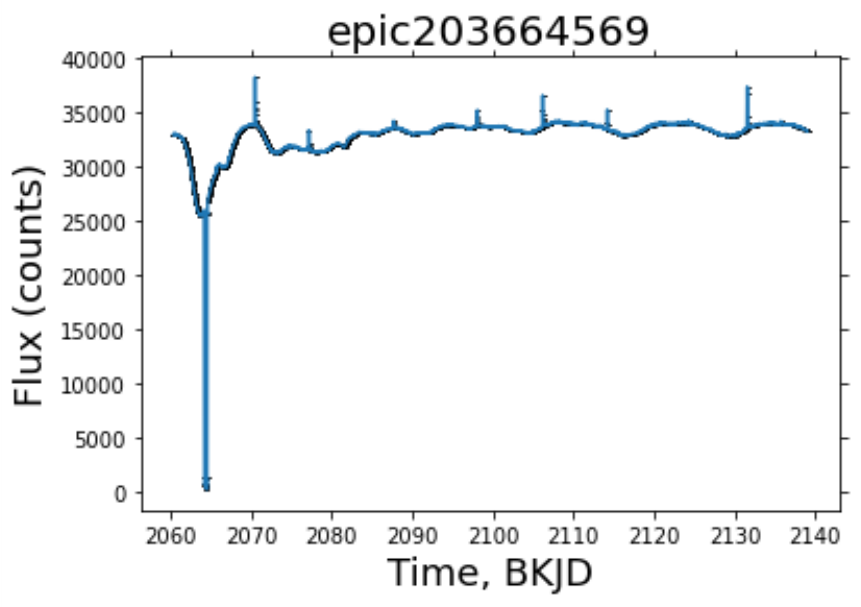}
    \includegraphics[width=.34\textwidth, height=.1875\textwidth]{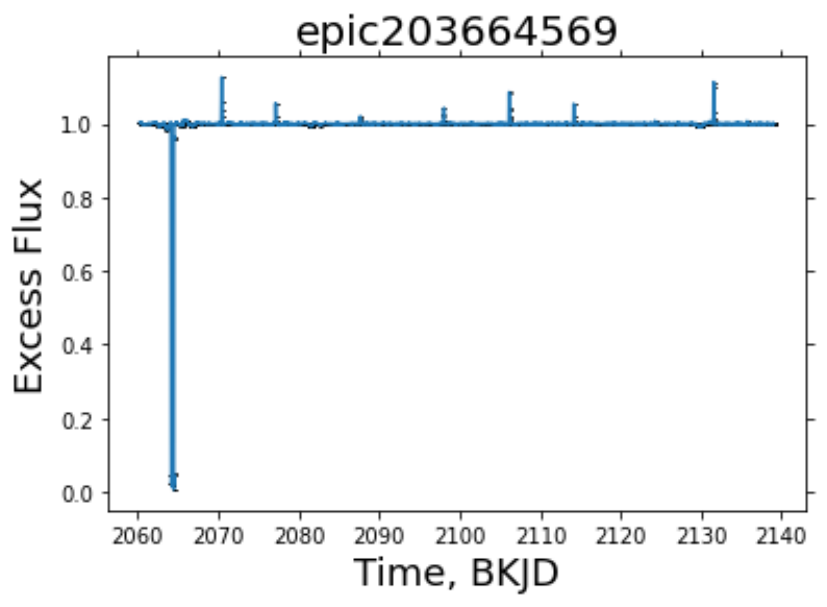}
    \includegraphics[width=.34\textwidth, height=.1875\textwidth]{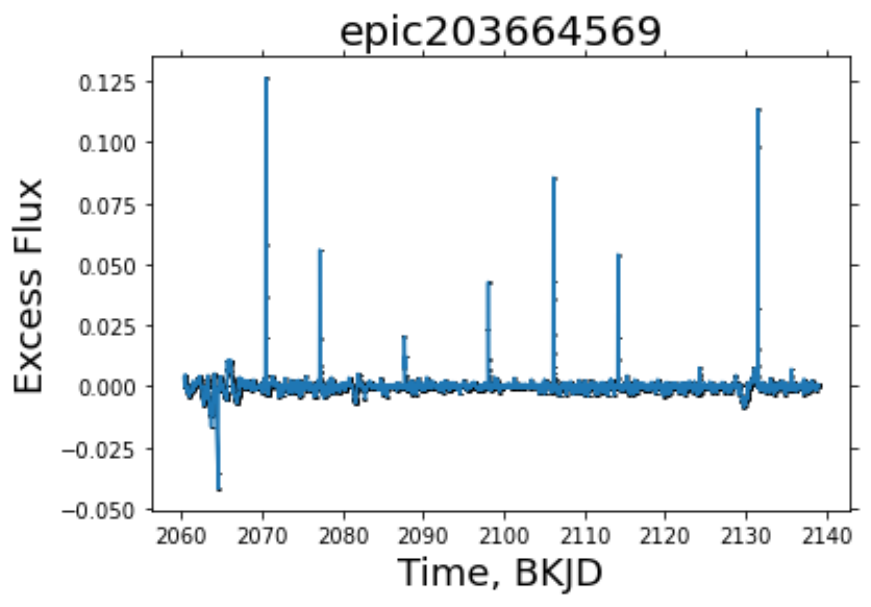}
    \includegraphics[width=.34\textwidth, height=.1875\textwidth]{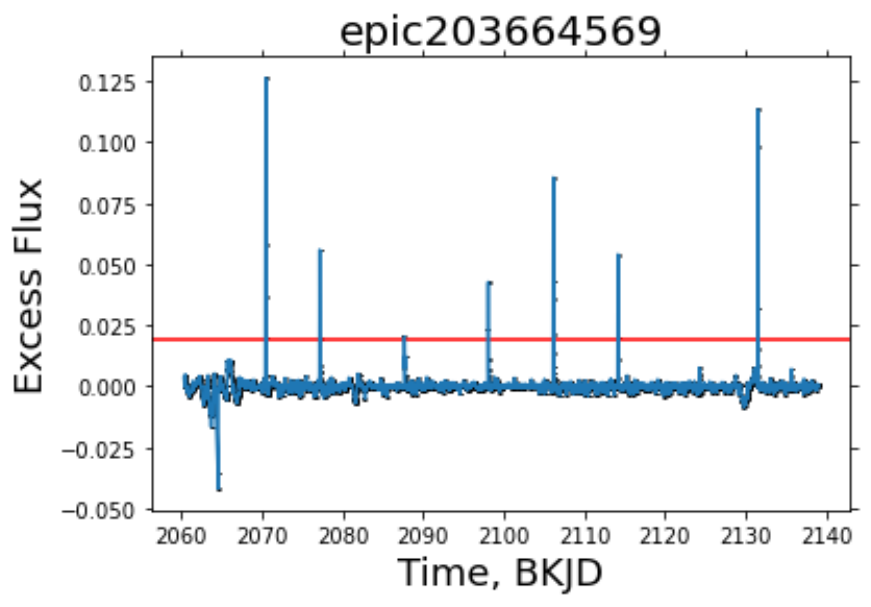}
    \caption{Full data of EPIC 203664569: Each of the panels above show the flux observed from EPIC 203664569.
    Upper Left: The full, unaltered lightcurve.
    Upper Right: The full lightcurve after flattening via the LightKurve flatten function.
    Lower Left: The flattened light curve after cutting outliers and shifting the average value to 0 rather than 1.
    Lower Right: Same as lower left with an indicator for the 4$\sigma$ threshold above which brightness peaks are collected for further study. Time presented in BKJD, the Barycentric Kepler Julian Date.}
    \label{fig:1-all}
\end{figure*}

We begin the process of working with these data by taking the raw flux measurements and passing them into the LightKurve python package to smooth modulations and leave us with a flattened light curve \citep{LK}. This is shown in the top two panels of Figure~\ref{fig:1-all} for EPIC 203664569. We then subtract the median from the light curve and apply a cut, masking any outliers above $5\sigma$ and below $-4\sigma$. With the outliers taken care of, we then define a threshold above which potential flare events are counted and isolated for further examination. The threshold of flare detection can be arbitrary, but is often set as $3\sigma$ \citep{rivera2023,mossoux2017}. Our threshold is taken as $4\sigma$, as bight flares are more likely to produce detectable echoes, and is shown by the red horizontal line in the last panel of Figure~\ref{fig:1-all}.

\begin{figure*}[!ht]
    \centering
    \includegraphics[width=.3\textwidth, height=.18\textwidth]{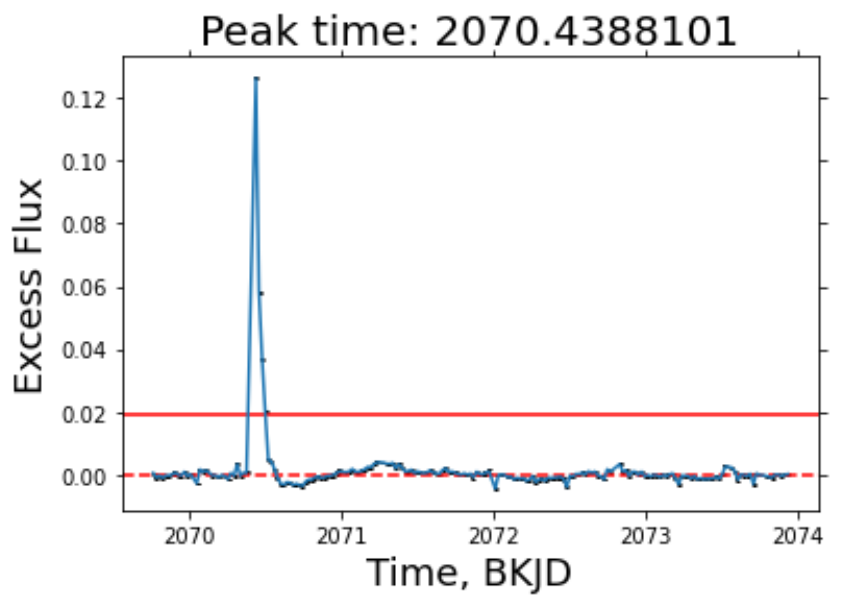}
    \includegraphics[width=.3\textwidth, height=.18\textwidth]{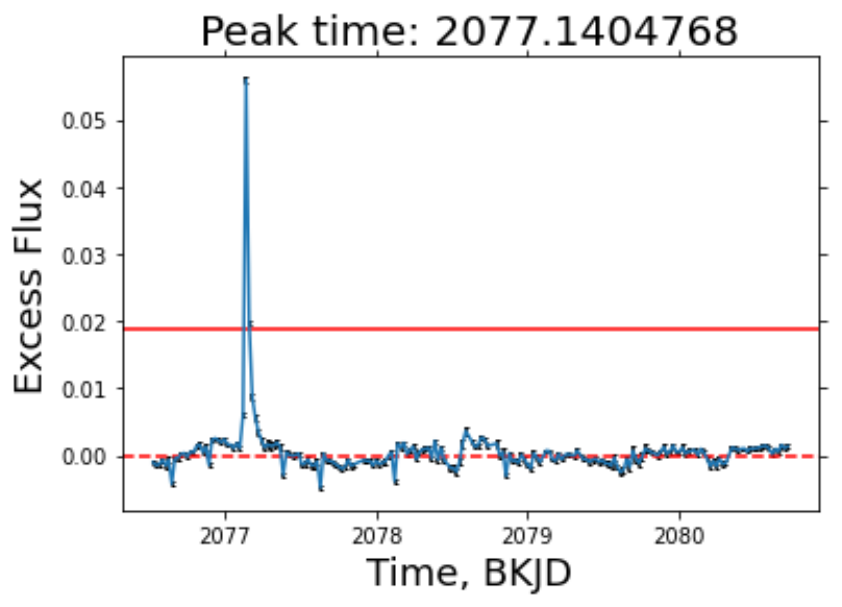}
    \includegraphics[width=.3\textwidth, height=.18\textwidth]{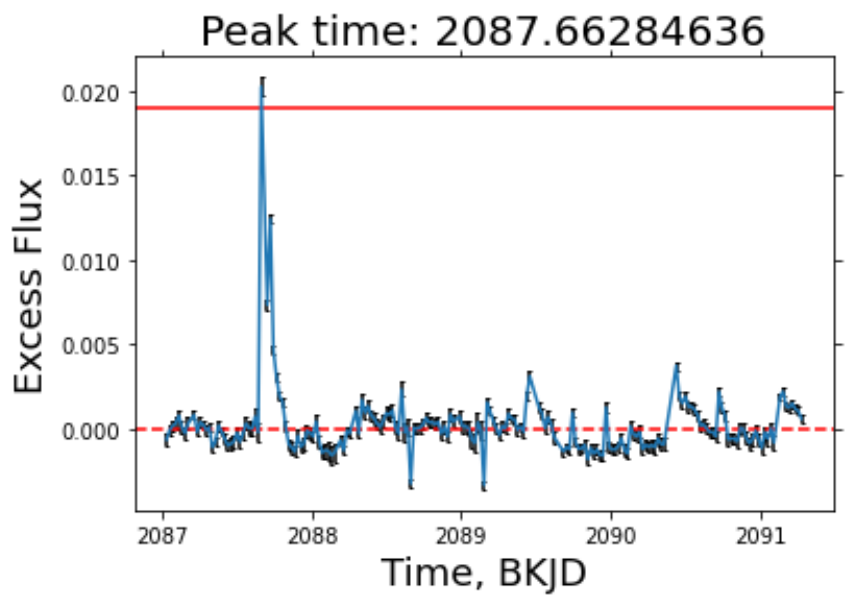}
    \includegraphics[width=.3\textwidth, height=.18\textwidth]{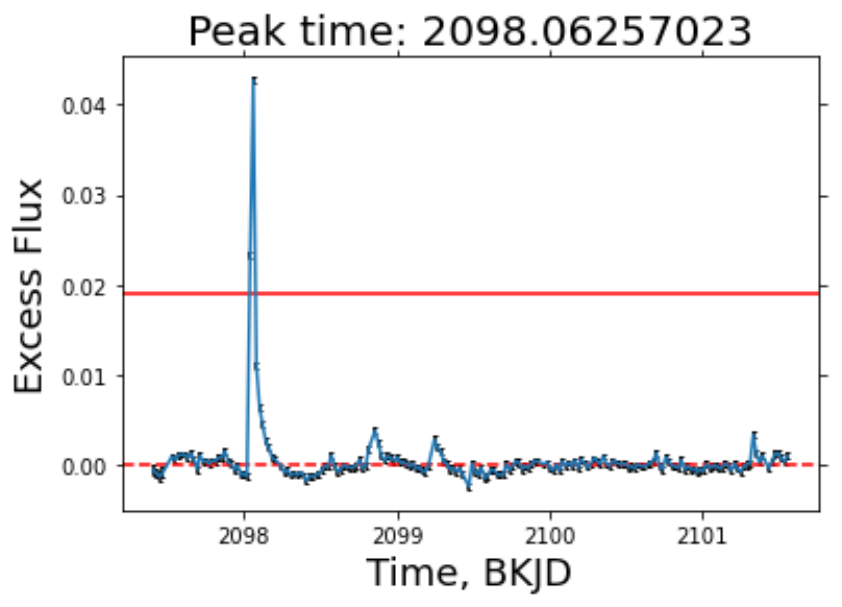}
    \includegraphics[width=.3\textwidth, height=.18\textwidth]{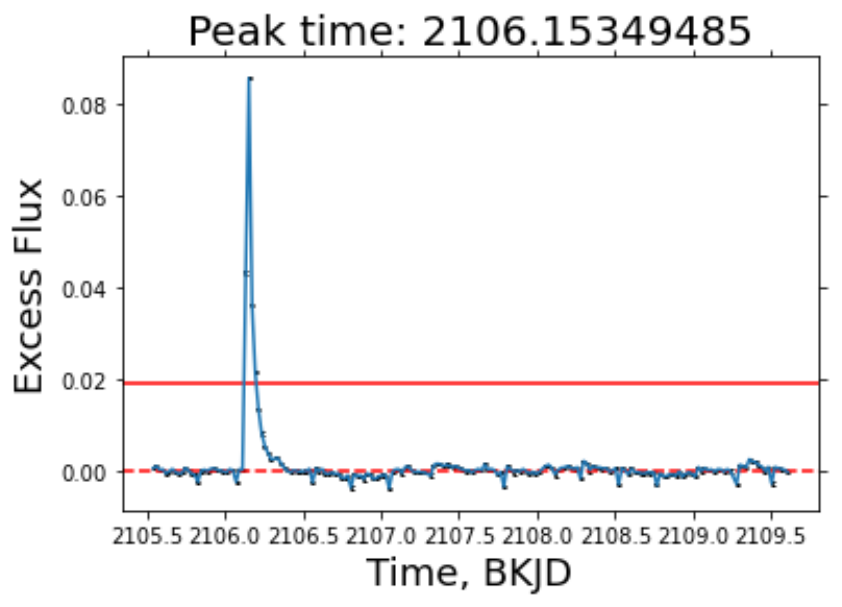}
    \includegraphics[width=.3\textwidth, height=.18\textwidth]{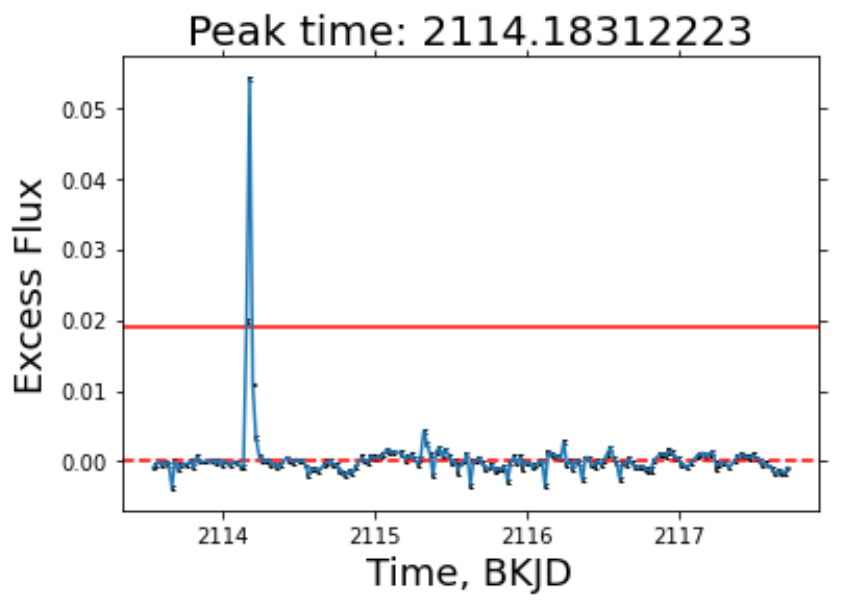}
    \includegraphics[width=.3\textwidth, height=.18\textwidth]{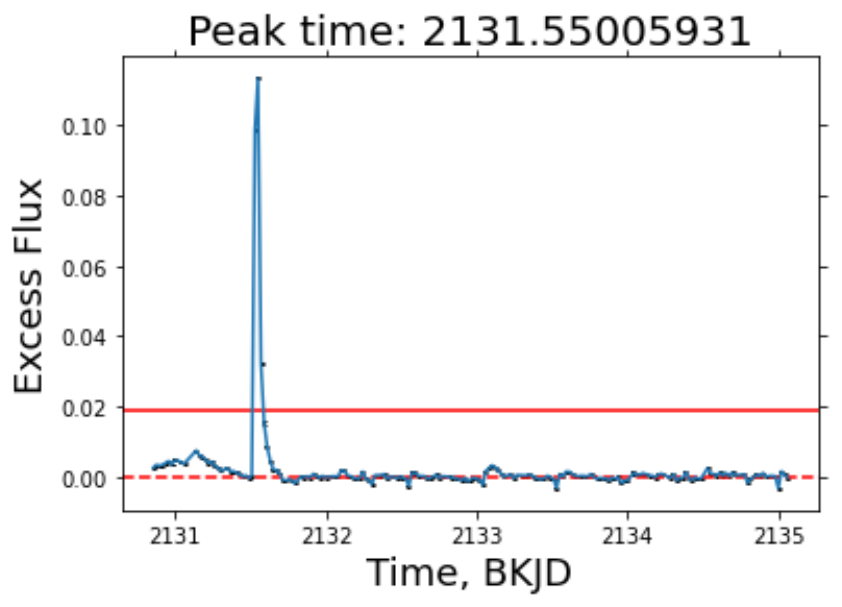}
    
    \caption{Peaks of EPIC 203664569: These 7 panels show a brightness peak at a specific BKJD time and the following light curve. Each of these continues the display of the flare threshold cut (red horizontal line) show in the 4th panel of Figure~\ref{fig:1-all}}
    \label{fig:1-peaks}
\end{figure*}

With our potential flare events identified, we isolate and plot them individually. These individual plots take about 10 time bins prior to the peak of the event to show the structure of the rise in brightness. After the peak, we display approximately 3 days worth of following flux. This is shown in Figure~\ref{fig:1-peaks}

\begin{figure*}[!ht]
    \centering
    \includegraphics[width=.37\textwidth, height=.25\textwidth]{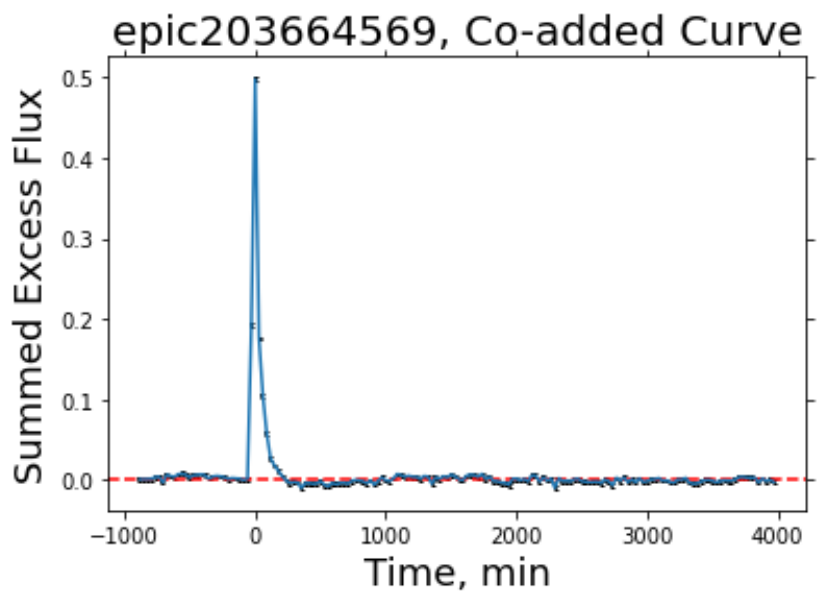}
    \includegraphics[width=.4\textwidth, height=.25\textwidth]{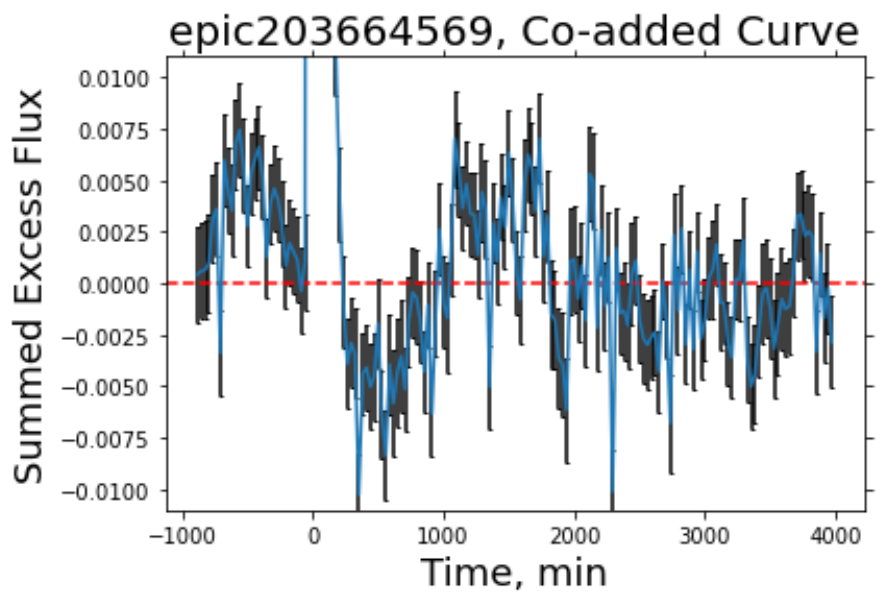}
    
    \caption{Co-added curves of EPIC 203664569 peaks: The left panel shows the result of coadding the peaks and their following lightcurves of Figure~\ref{fig:1-peaks}. The right panel shows a scaled version of the left panel to better highlight the low-level excess on either side of the peak. The post-flare data show some excess brightness between 1000 and 1700 minutes, possibly indicative of a faint echo. We note that we see similar excess on the pre-peak side as well, possibly indicative of pre-flare brightness increases.}
    \label{fig:1-summed}
\end{figure*}

We then take these isolated potential flare events and shift in time to align the flare peaks and then co-add the light curves. This process follows with those described in \cite{bromley2021, King24, King24-2}, which state that co-adding multiple flare events can yield a stronger, more readable echo signal. We show this full co-added curve in Figure~\ref{fig:1-summed}, along with the portion of the lightcurve from 500-4000 minutes post-peak. The post-peak curve shows a hint of excess brightness between 1000 and 1700 minutes. This structure is possibly indicative of a faint echo. Summing the flux of this segment yields a value of about 0.085. This is about $10\%$ of the flux produced by first five time bins of the co-added curve. We note the presence of pre-peak flux excess of similar magnitude to the post-peak segment of interest. This feature appears to arise primarily from the curve shown in panel 7 of Figure~\ref{fig:1-all}, where a clear ``bump" of flux is visible in the pre-peak curve.

With this segment of data prepared, we apply the same MCMC fitting process as described in Section \ref{subsec:mcmc}. Disk radii are allowed to vary greatly compared to the ranges defined in \cite{King24-2}. This is due to the switch from short-cadence modeling to long-cadence, where probing much larger distances is possible. We allow $r_{\rm in}$ walkers to roam between 10-500 AU. We cap $r_{\rm out}$ at 600 AU to keep model echoes closer to the timescale shown in our data. Additionally, we constrain outer radius values to those greater than their corresponding inner radius. Inclination is held between 0 and 90 degrees and albedo is allowed to vary from 0 to 1. 

\begin{figure*}[!ht]
    \centering
    \includegraphics[width=.7\textwidth, height=.5\textwidth]{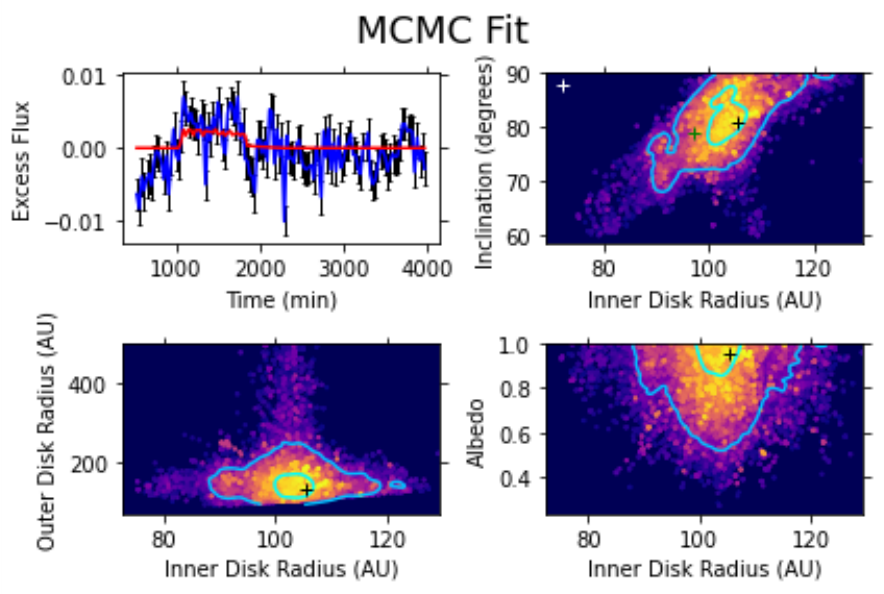}
    
    \caption{MCMC fit of EPIC 203664569:
    Top Left: This panel shows the ideal model echo (red) produced by the MAP parameters returned from the MCMC fit overlayed on the co-added data (blue).
    Remaining panels each show posterior parameter distributions with blue and cyan contours depicting $1 \sigma$ and $3\sigma$ confidence ranges, respectively. The black plus signs indicate the most probable parameters in each space. The white plus sign in the top-right panel shows the parameters for inclination and inner radius reported by \cite{barenfeld17}. The green plus shows the lower limit of inclination and upper limit of inner radius reported by \cite{barenfeld17}.}
    \label{fig:1-fit}
\end{figure*}

The resulting MCMC fit produces the parameter-space plots shown in Figure~\ref{fig:1-fit}. The model predicts that the disk around this young star has parameters with $1\sigma$ uncertainties of $r_{\rm in}=105\ (-5,+1) $AU, $r_{\rm out}=130 (-9,+29) $AU, $i=81^\circ \ (-2,+5)$, and $\gamma=0.95 \ (-.10,+0.03)$. We compare these values with those reported by \cite{barenfeld17}. Here, EPIC 203664569 (reported as 2MASS J16163345-2521505) is found to have an inner dust radius of 72 (-23,+25) AU and an inclination of 88 (-9,+2) degrees. We find strong agreement in the inclination calculation, which is very encouraging. The lower range on our calculated inner radius nearly overlaps within the the upper range reported by \cite{barenfeld17}. This is a highly encouraging result at this stage.

After running our MCMC routine with the cylindrical wall geometry, the model failed to meaningfully constrain parameters or fit the given light curve. From this we conclude that a cylindrical geometry is not appropriate for fitting this system.

\subsection{EPIC 247584113}

\begin{figure*}[!ht]
    \centering
    \includegraphics[width=.34\textwidth, height=.2\textwidth]{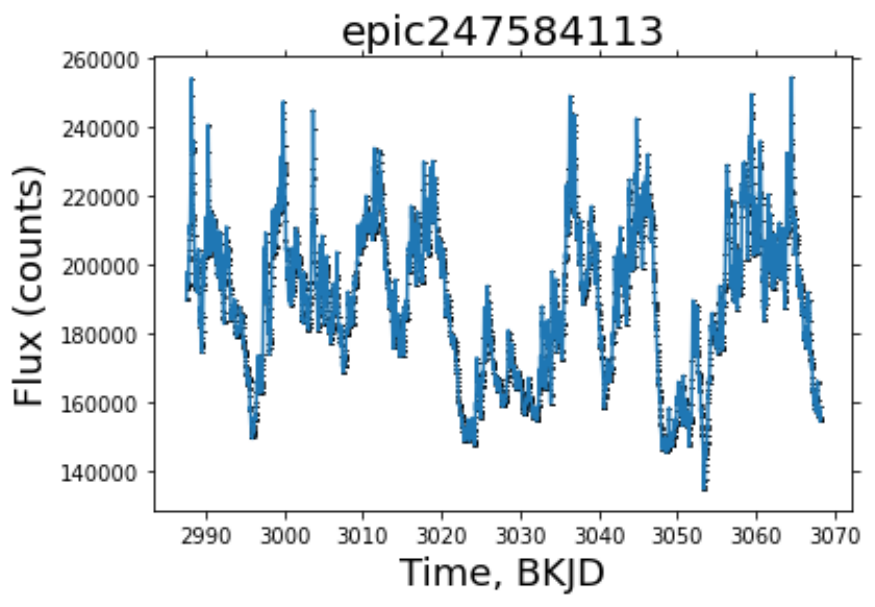}
    \includegraphics[width=.34\textwidth, height=.2\textwidth]{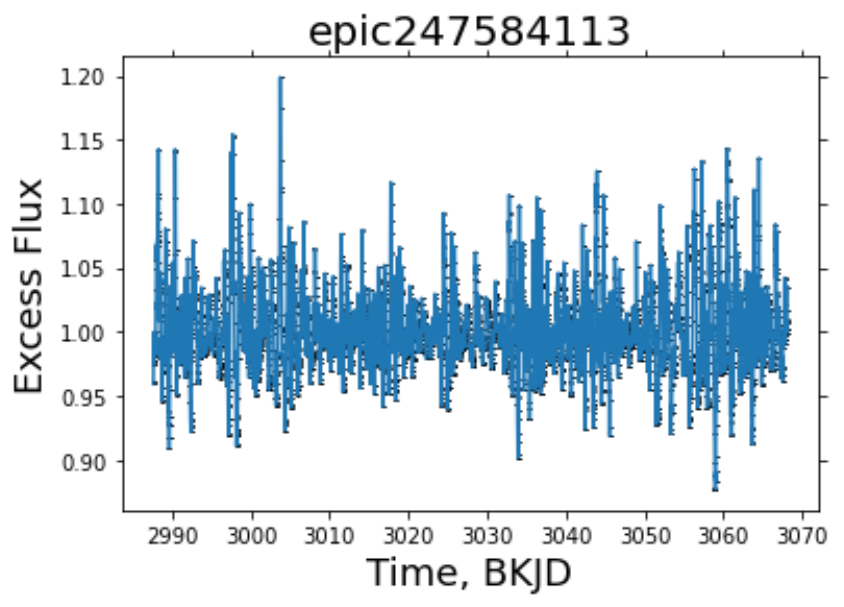}
    \includegraphics[width=.34\textwidth, height=.2\textwidth]{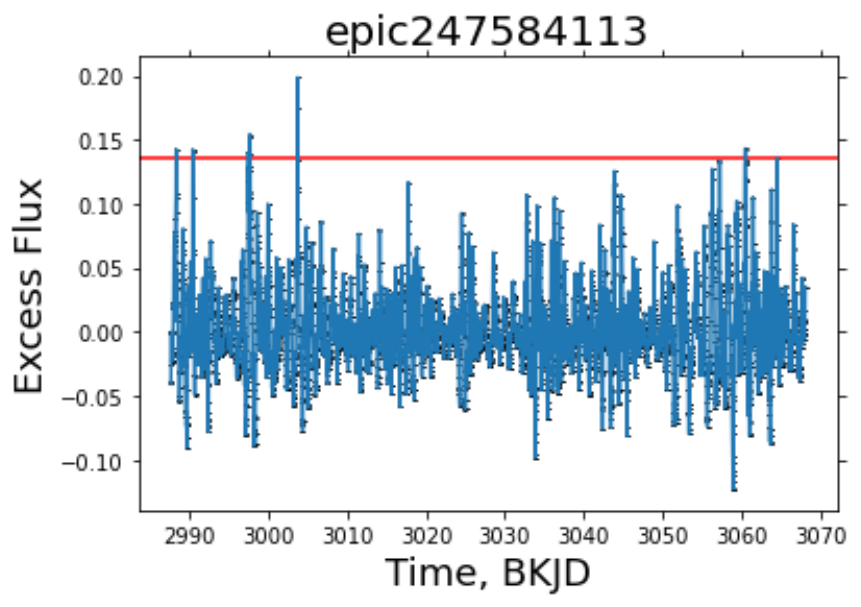}
    \caption{Full Data on EPIC 247584113: Each of the panels above show the flux observed from EPIC 247584113.
    Upper Left: The full, unaltered lightcurve.
    Upper Right: The full lightcurve after flattening via the LightKurve flatten function.
    Lower: The flattened light curve after cutting outliers and shifting the average value to 0 rather than 1 and an indicator for the 4$\sigma$ threshold above which brightness peaks are collected for further study.}
    \label{fig:2-all}
\end{figure*}

\begin{figure*}[!ht]
    \centering
    \includegraphics[width=.3\textwidth, height=.18\textwidth]{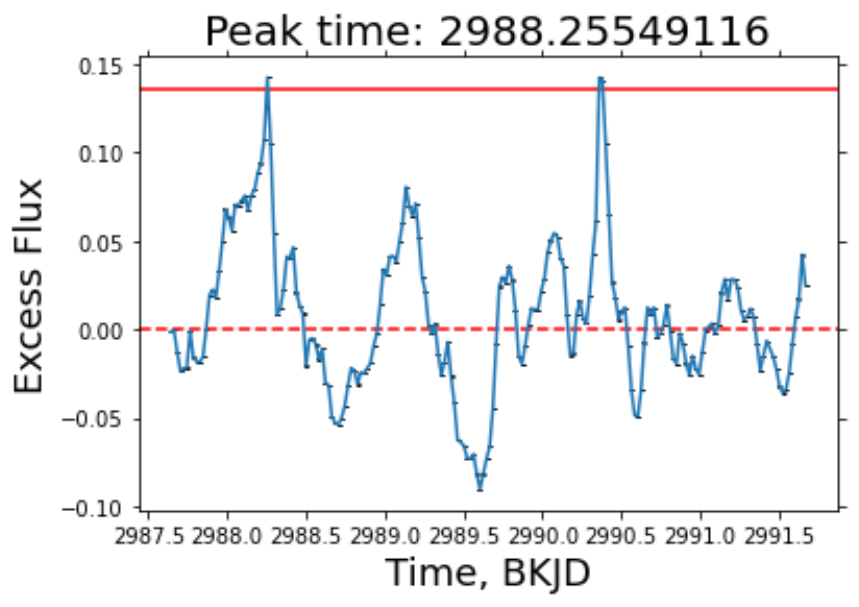}
    \includegraphics[width=.3\textwidth, height=.18\textwidth]{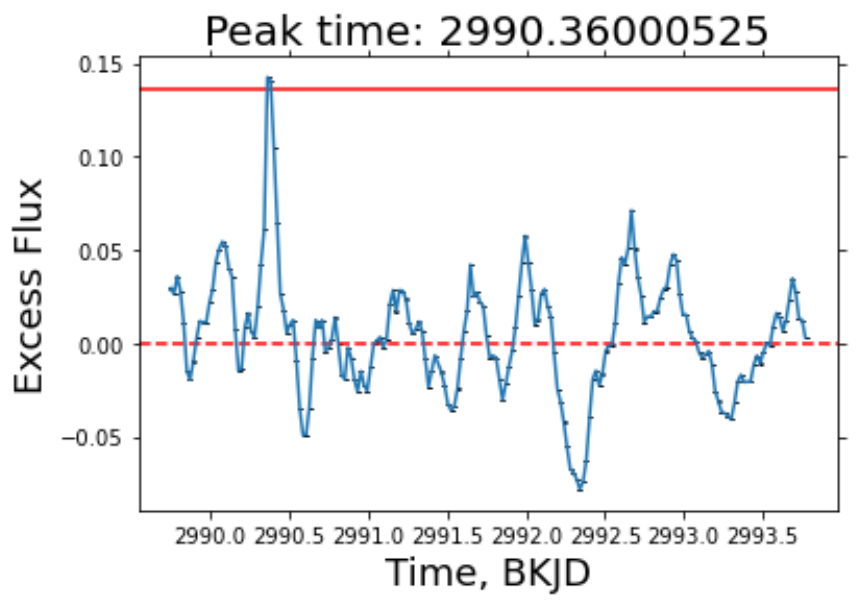}
    \includegraphics[width=.3\textwidth, height=.18\textwidth]{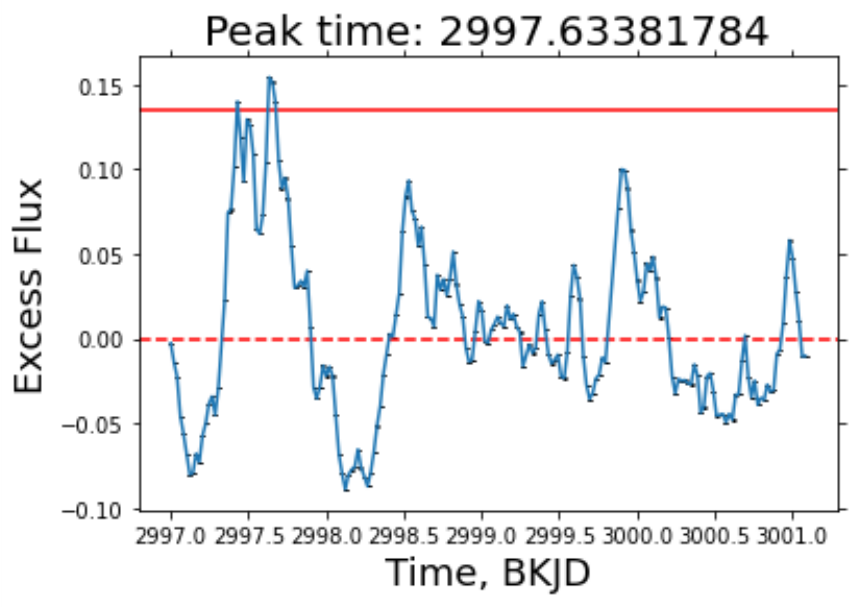}
    \includegraphics[width=.3\textwidth, height=.18\textwidth]{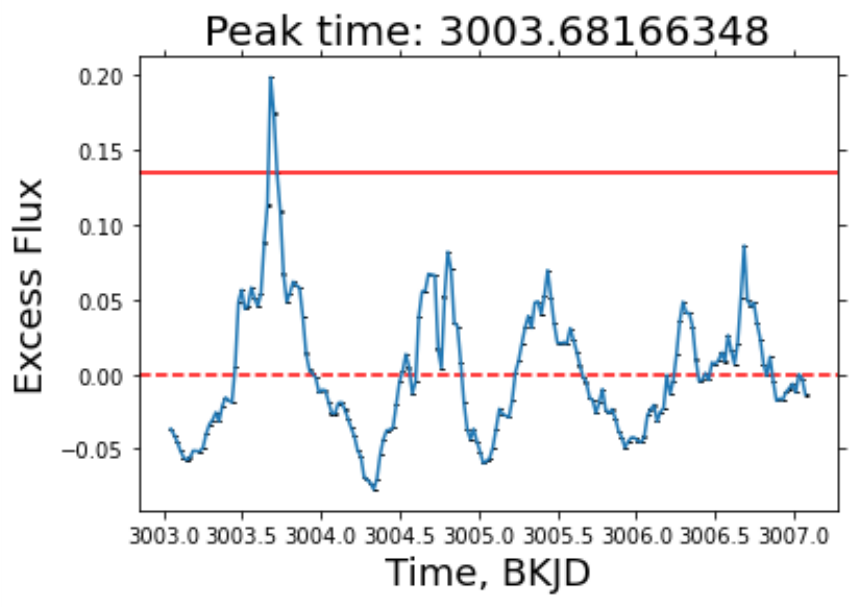}
    \includegraphics[width=.3\textwidth, height=.18\textwidth]{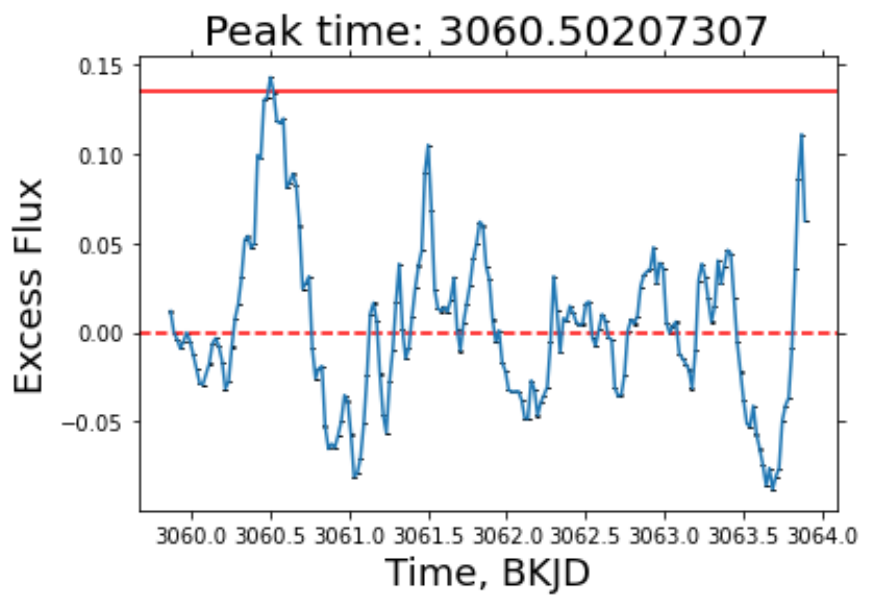}
    \includegraphics[width=.3\textwidth, height=.18\textwidth]{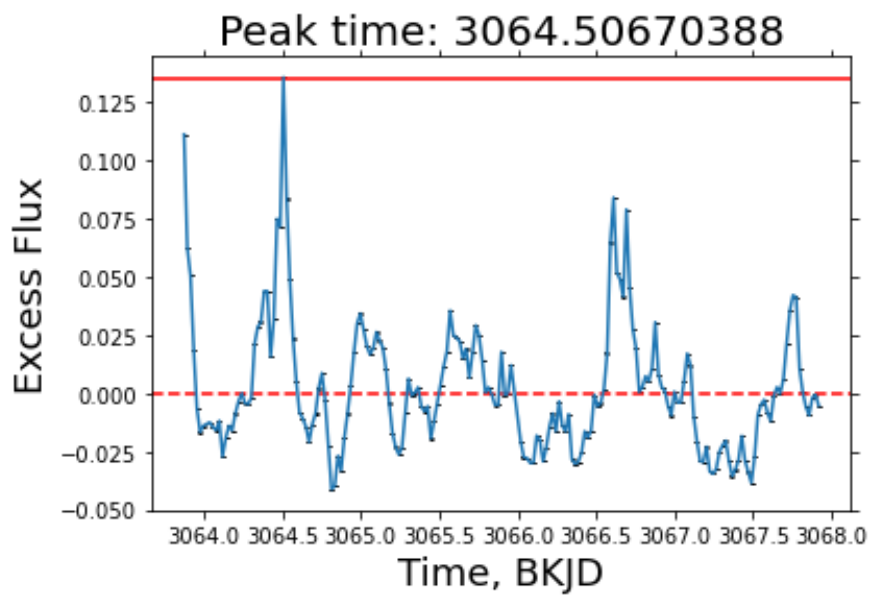}
    
    \caption{Peaks of EPIC 247584113: These 6 panels show a brightness peak at a specific BKJD time and the following light curve. Each of these continues the display of the flare threshold cut (red horizontal line) shown in the 3rd panel of Figure~\ref{fig:2-all}. We note that the peak shown in the second panel is also present in the first panel, there is much more modulation throughout these figures compared to those in Figure~\ref{fig:1-peaks}, and most of these events do not show a characteristic jump and gradual decline seen typically in flares.}
    \label{fig:2-peaks}
\end{figure*}

We now present a star which does not appear to be a compelling candidate for the light echo method. Figure~\ref{fig:2-all} presents light curves of EPIC 247584113, collected and processed in the same manner as those shown in Figure~\ref{fig:1-all}. Our $4\sigma$ threshold for peak brightness events yields 6 peaks, shown in Figure~\ref{fig:2-peaks}. To put it plainly, these results are far less `clean' than those shown in Figure~\ref{fig:1-peaks}. We see one instance of a peak event occurring very soon after another (Panel 1, with the second peak shown again in Panel 2). There is significantly more noticeable modulation in the post-peak curves compared to Figure~\ref{fig:1-peaks} and the peaks themselves mostly do not present the structure expected from flare-like events. These may instead be the result of stellar brightness fluctuations, which could also explain the greater variation in the post-peak curves. 

\begin{figure*}[!ht]
    \centering
    \includegraphics[width=.4\textwidth, height=.25\textwidth]{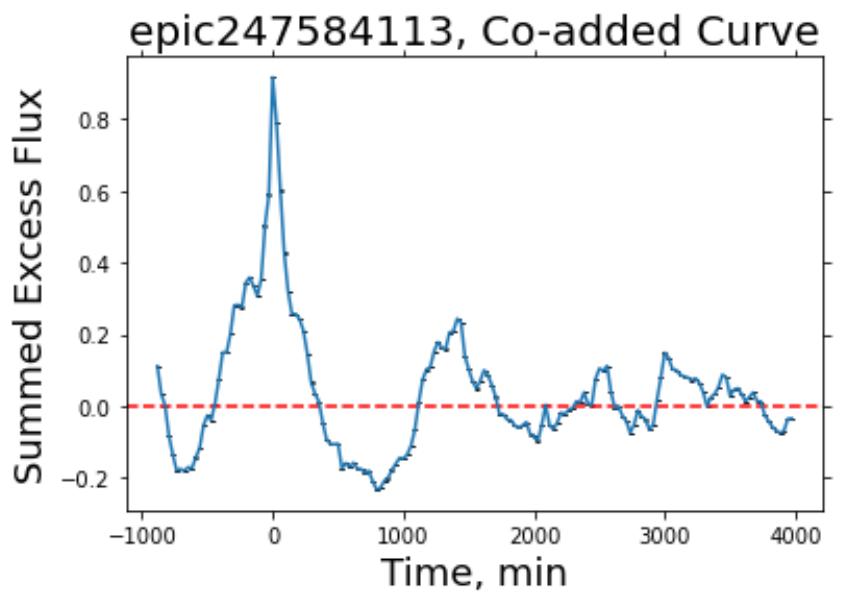}
    
    \caption{Co-added curves of EPIC 247584113 peaks: This figure shows the result of coadding the peaks and their following lightcurves of the plots from Figure~\ref{fig:2-peaks}. The post-flare data show significant modulation that cannot be fit by our light echo model.}
    \label{fig:2-summed}
\end{figure*}

We further explore this object by summing the post-peak curves in the same fashion as done for the previous object. Figure~\ref{fig:2-summed} shows the result of this co-addition process. The second panel of this figure shows only the post-peak light curve. 

Unlike Figure~\ref{fig:1-summed}, a zoomed-in post-peak panel is not necessary to see the post-peak structure, which is clearly visible in Figure~\ref{fig:2-summed} as well. This structure is far too bright to be the result of an echo produced in our model. In fact, our MCMC fitting routine described previously fails to fit this curve at all, and thus we do not have a figure corresponding to Figure~\ref{fig:1-fit} for this object. We determine that while there is some interesting structure here, any brightness fluctuations following a peak event from this star are likely further intrinsic modulation of the star's brightness itself and not the result of an echo.

\section{Summary}\label{sec:summary}

Here we presented an updated model of the light echo method described in \citet{King24-2}, capable of handling realistic, time-resolved flares. We demonstrated the necessity of this change using an MCMC fitting routine, as modeling with delta function flares could not accurately estimate disk parameters with an echo produced by time-resolved flares.

We tested the light echo model of time-resolved flares on Kepler data and found that EPIC 203664569 is a great candidate for further exploration with this method. Parameter estimation from fitting with our model yields an inner disk radius of $r_{\rm in}=105\ (-5,+1) $AU, an outer radius of $r_{\rm out}=130 (-9,+29) $AU, an inclination of $i=81^\circ \ (-2,+5)$, and albedo $\gamma=0.95 \ (-.10,+0.03)$. These values show consistencies with those reported by \cite{barenfeld17} and are a very encouraging sign for the potential of this method to extract disk parameters from post-flare light curves. We predict that further observations of flare events of this star could yield sufficient data to confirm this prediction. 

In contrast, EPIC 247584113 was shown to be not a viable candidate for this model due to its more variable nature. We predict that further observations of this star would show similar post-peak variability, but likely with differing structure and at different post-peak times, ruling out the possibility of light echoes as their origin.

It is important to note that we are dealing with very small flux values in this work and that flares can certainly exhibit unexpected post-peak characteristics. Flares also illuminate only the half of the disk they face. Future work is required with many more observations of flare events to help reduce potential effects of both subflaring and flare location on the anticipated signal of light echoes. An additional means for distinguishing scattered flare light from sub-flare structure could be polarization, as discussed by \cite{King24}, though this also requires many observations to yield a sufficient signal. Future work could also benefit from use of cross-correlation, or photo-reverberation mapping, as an additional means of differentiating echo signatures from stellar activity \citep{Mann, Sparks2018}.

We are excited about the progress of the light echo method as showcased by this work. The results presented here demonstrate its promise for extracting disk parameters from post-peak light curves, allowing us to better study the structure of young planetary systems and, eventually, discover unresolved disks around young stars.

\section{Acknowledgments}

We thank Dr. Ann Marie Cody and Dr. Peter Plavchan for their conversations regarding the content of this work and the data they directed us to from \cite{cody18,venuti21,cody22}.

We also thank an anonymous referee for their insight and suggestions to improve this paper.

\bibliographystyle{aasjournal.bst}
\bibliography{bib}

\end{document}